\documentclass[%
 reprint,
 amsmath,amssymb,
 aps,
]{revtex4}

\usepackage{floatrow}
\usepackage{verbatim}
\usepackage{graphicx}
\usepackage{dcolumn}
\usepackage{bm}
\usepackage[position=t,singlelinecheck=off]{subfig}
\usepackage{caption}

\begin{document}

\preprint{APS/123-QED}

\title{Thermal conductivity of h-BN monolayers using machine learning interatomic potential}

\author{Yixuan Zhang}

\author{Chen Shen}
\email{chenshen@tmm.tu-darmstadt.de}
\author{Teng Long}

\author{Hongbin Zhang}
 \email{hongbin.zhang@tu-darmstadt.de}
\affiliation{
 Institute of Materials Science, Technical University of Darmstadt, Darmstadt 64287, Germany
}

\date{\today}

\begin{abstract}
Thermal management materials are of critical importance for engineering miniaturized electronic devices, 
where theoretical design of such materials demands the evaluation of thermal conductivities which are numerically expensive. 
In this work, we applied the recently developed machine learning interatomic potential (MLIP) to evaluate the thermal conductivity of hexagonal boron nitride monolayers. The MLIP is obtained using the Gaussian approximation potential (GAP) method, and the resulting lattice dynamical properties and thermal conductivity are compared with those obtained from explicit frozen phonon calculations. It is observed that accurate thermal conductivity can be obtained based on MLIP constructed with about 30\% representative configurations, and the high-order force constants provide a more reliable benchmark on the quality of MLIP than the harmonic approximation. 
\end{abstract}

\maketitle


\section{\label{sec:Introduction}Introduction}

Hexagonal boron nitride (h-BN), as one of the most interesting III-V compounds with  layered structures, has drawn intensive attentions in the last few decades~\cite{Geick:1966,Sichel:1976,Kubota:2007,Lin:2014,Tran:2016,Jiang:2018}, as h-BN exhibits many intriguing properties such as excellent mechanical properties~\cite{Kiran:2011}, high in-plane thermal conductivity~\cite{Lin:2014}, 
and behaves like ionic half-metal~\cite{Kim:2011,Zheng:2008} due to the different electronegativity of the boron and nitrogen atoms.
Such properties make h-BN a promising thermal management material which is essential for engineering miniaturized electronic devices.
On the other hand, the thermal conductivities play also a crucial role in optimizing the thermoelectric performance~\cite{Chiu:2016}.
Previous theoretical studies predicted an in-plane thermal conductivity as high as 550 W/mK~\cite{Jiang:2018} at room temperature, consistent with the reported experimental values ranging from 220 to 420 W/mK~\cite{Sichel:1976,Jiang:2018,Simpson:1971}. 
The evaluation of thermal conductivities requires accurate calculations on phonon-phonon interaction, where at least third-order force constants have to be calculated. For instance, such calculations of thermal conductivities are usually done using the frozen phonon method based on density functional theory (DFT)~\cite{Fu:1985,Frank:1995}, which are numerically demanding.

Recently, machine learning interatomic potential (MLIP) has become a valuable solution to construct atomistic models with dominant short-range interactions. For instance, Gaussian approximation potential (GAP) introduced by Bartók {\it et al.}~\cite{Bartok:2010,Szlachta:2014} has been applied to describe the potential energy surface (PES) with sufficient accuracy, where the interatomic potential (energy, forces, and strain) are fitted using the Gaussian kernel method~\cite{Rasmussen:2004,Seeger:2004} on a multidimensional configurational space of atomic positions. Szlachta {\it et al.} established a comprehensive GAP model based on a database of bcc tungsten and applied this model to calculate the properties such as screw dislocation, the Peierls barrier and vacancy-dislocation interaction energies~\cite{Szlachta:2014}. John and Csányi introduced a framework describing general many-body coarse-grained interactions of molecules. By combining molecular dynamics (MD) and GAP potential, a decent model which can precisely describing the free energy surface of molecular liquids were established~\cite{John:2017}. Qian and Yang developed two GAP models for both hcp and bcc phase of zirconium crystals, and used 
them in MD calculations to predict the phonon dispersion at finite temperature~\cite{Qian:2018}. 
Recently, Deringer {\it et al.} gave a brief introduction of MLIP and highlighted some representative applications of that in material science, {\it e.g.}, the prediction of crack behavior of Si, the simulation of phase-change materials, the calculation of nanoparticles for catalysis and the investigation of carbon-based nanomaterials ~\cite{Deringer:2019}.

In this work, we demonstrated a MLIP obtained using GAP which can be applied to accurately evaluate  the thermal conductivity of h-BN monolayer. Detailed comparisons on the phonon dispersion, density of states, phonon lifetime, and thermal conductivities are done between the explicit frozen phonon and MLIP calculations. Particularly, we studied how the numerical efforts can be reduced while keeping the accuracy via testing the minimum size of dataset for the GAP model. This establishes a workflow combining DFT and ML to evaluate the thermal conductivities efficiently, which paves the way for future design of materials for thermal management and thermoelectric applications.

\section{\label{sec:NumericalMethod}Numerical Methods}

To evaluate the phonon and phonon-phonon interactions, second- and third-order force constants are calculated using the Alamode package~\cite{Tadano:2014}. The configurations with different displacement patterns are generated using the frozen phonon method. The atomic forces for each atoms and total energies of these displaced configurations are calculated by DFT, performed using the projector augmented-wave method as implemented in Vienna Ab-Initio Simulation Package (VASP)~\cite{Kresse:1996a,Kresse:1996b}. The energy cutoff is $550$ eV and the energy convergence tolerance is $10^{-8}$ eV.
 The exchange correlation functional is chosen to be the generalized gradient approximation parametrized by Perdew-Burke-Ernzerhof (GGA-PBE)~\cite{Perdew:1996}. A supercell of $6\times6\times1$ is constructed, with the interlayer distance being $20$ \AA. The Brillouin zone is sampled using the gamma centered Mokhorst-Pack scheme with a $4\times4\times1$ k-mesh with Bloch corrections~\cite{Blochl:1994}.

The thermal conductivity is evaluted based on the Boltzmann transport theory with the relaxation time approximation (BTE-RTA)~\cite{McGaughey:2004}, which yields

\begin{equation}
\kappa_{ph}^{\mu\upsilon}(T)=\frac{1}{\Omega N_q}\sum_{q,j}C_{q,j}(T)v_{qj}^{\mu}v_{qj}^{\upsilon}\tau_{qj}(T) \label{eq:equation1}
\end{equation}

where $\Omega$ is the volume of the unit cell and $N_q$ is the number of q-point, $C_{q,j}(T)=\hbar\omega_{qj}\partial{n_{qj}}/\partial{T}$ is the constant-volume specific heat with $n_{qj}=1/(e^{\hbar\omega_{qj}/kT}-1)$ being the Bose-Einstein distribution function, $v_{qj}^{\mu/\upsilon}$ denotes the group velocities along the Cartesian $\mu/\upsilon$-direction, $\tau_{qj}(T)$ is the phonon lifetime which can be estimated according to the Matthiessen’s rule as $\tau_{qj}(T)=\frac{1}{2}(\Gamma_{qj}(T))^{-1}$ with $\Gamma_{qj}(T)$ is the imaginary part of the phonon self-energy considering the lowest-order three-phonon interaction.

The use of $6\times6\times1$ supercell results in 867 3rd order frozen phonon configurations, where the atomic displacement amplitude is set to 0.04 \AA. The total energies and forces of such supercells are calculated and fed into the GAP~\cite{Bartok:2010} to construct MLIP. It is noted that due to the 2D nature of BN monolayers, the lattice dynamics along the z-direction is different from the in-plane directions, thus we performed a classification of frozen-phonon configurations based on the direction of atomic displacement. That is, 867 configurations are separated into two groups: one group (dubbed database1 with 554 configurations) with only in-plane displacements and the other group (dubbed database2 with 313 configurations) with z-direction displacement. To ensure the diversity and continuity of the training sets, and to develop a thorough understanding on how to pick up representative configurations as well, we propose a general but robust scheme (denoted as “Manual”) for selecting training set. We consider the configurations with displacements on the same atom but of opposite directions as similar configurations, such configurations are grouped as subsets for database1 and database2. The training sets are constructed by choosing one configuration from each subsets, leading to 160 out of 554 configurations from database1 and 165 out of 313 configurations from database2.

Moreover, to estimate the minimum number requirement of configurations for accurately reproducing the thermal conductivity, and also the feasibility of employing active learning to make this workflow automatically, another selecting method (denoted as “Incremental”) was tested, where the most representative configurations are selected out as the training set with the help of MLIP. Specifically, starting from a random configuration for each database, at each iteration, 25 (15) configurations for database1 (database2) were screened out and added into the training set for studying the model for the next iteration. Then the results of different iteration steps (2, 3, 4, 5, 6) were compared. The selected training set with the corresponding percentage are shown in Table \ref{tab:table1} for different selecting methods. Note that the crystalline symmetry is not considered here, in order to test different ways to obtain the training sets.

\begin{table}
\caption{\label{tab:table1}The number of the training sets of manual and different incremental method and their relative proportion out of the database. 
}
\begin{ruledtabular}
\begin{tabular}{lcc}
\textrm{}&
\textrm{Training set}&
\textrm{Proportion(\%)}\\
\colrule
Manual & 325 & 37.5\\
Incremental-2 & 82 & 9.5\\
Incremental-3 & 122 & 14.1\\
Incremental-4 & 162 & 18.7\\
Incremental-5 & 202 & 23.3\\
Incremental-6 & 242 & 27.9\\
\end{tabular}
\end{ruledtabular}
\end{table}

The construction of GAP is performed using QUIP package \cite{Bartok:2015}. In GAP, the total energy of MLIP consists of the local energies of each individual atoms, and the covariance of two local energies is built in the forms of kernel function:
\begin{equation}
E=\sum_i\varepsilon(\bm{q}_i) \label{eq:equation2}
\end{equation}

\begin{equation}
\langle\varepsilon(\bm{q}_i)\varepsilon(\bm{q}_j)\rangle=\alpha K(\bm{q}_i,\bm{q}_j) 
\label{eq:equation3}
\end{equation}

where $E$ is the total energy of the system, $\varepsilon(\bm{q}_i)$ and $\varepsilon(\bm{q}_j)$ are the energy contribution from atom $i$ and $j$, $\bm q$ is a descriptor whose function is to characterize the atomic environment, {\it e.g.}, positions of atoms, in the neighborhood of the target atom. $\bm{q}_i$ here can be considered as the descriptor of atom i of the configurations to be predicted and $\bm{q}_j$ of atom j of the configurations in the training set. $K(q_i,q_j)$ is the kernel function which describes the similarity of atomic environment between the atom i and atom j. And $\alpha$ can be understood as the weight of this kernel, which can be determined by the fit. Then the interatomic forces can be calculated as the partial derivative of total energy versus displacement:

\begin{equation}
f_{k}=-\frac{\partial{E}}{\partial{r_{k}}}
\label{eq:equation4}
\end{equation}

where $r_{k}$ is the x, y or z component of the Cartesian coordinates of atom k. The weights are fitted by minimizing the lost function in kernel ridge regression, where the unknown function is expanded as a linear combination of kernel functions:

\begin{equation}
F(\bm{q})=\sum_i \alpha_i K(\bm{q},\bm{q}_i) 
\label{eq:equation5}
\end{equation}

\begin{equation}
L=\sum_i(t_i-F(\bm{q_i}))^2+\lambda\bm{\alpha}^T\bm{K}\bm{\alpha} 
\label{eq:equation6}
\end{equation}

where $t_i$ and $F(\bm{q_i})$ denote the observed and predicted total energy of system or interatomic force on atom, $\lambda$ is a hyperparameter and $\bm{\alpha}=(\alpha_1, \alpha_2, ...,\alpha_j, ...)$ is the vector of the weights to be fitted. $\bm{K}$ is a covariance matrix. The form of this matrix for energies and forces are shown respectively in Eq. \ref{eq:equation7} and Eq. \ref{eq:equation8}:

\begin{equation}
\bm{K}_{energy}=\langle E_N E_M\rangle=\alpha\sum_{i\in N} \sum_{j\in M}K(\bm{q_i},\bm{q}_j)
\label{eq:equation7}
\end{equation}

\begin{equation}
\bm{K}_{force}=\langle \frac{\partial{E_N}}{\partial{r_{k}}} E_M\rangle=  \frac{\partial \langle{E_N E_M}\rangle}{\partial{r_{k}}} =\alpha\sum_{i\in N} \sum_{j\in M} \bigtriangledown_{\bm{q}_i} K(\bm{q_i},\bm{q}_j) \frac{\partial{\bm{q}_i}}{\partial{r_{k}}}
\label{eq:equation8}
\end{equation}

In GAP, the spatial cutoff function is employed to ensure the computational efficiency. The implemented cutoff function here is:

\begin{equation}
f(x)=
\begin{cases}
1 &  \text{for $r \leq r_{cut}-d$} \\
\bigg[\cos\big(\pi\displaystyle{\frac{r-r_{cut}+d}{d}}\big)+1\bigg]/2 & \text{for $r_{cut}-d < r \leq r_{cut}$} \\
0 & \text{for $r > r_{cut}$}
\end{cases}
\end{equation}

with $d$ determines the width of the cutoff region. 

The descriptors used in this manuscript are pair and triplets descriptors. For pair descriptor, it is simply defined as the distances between the pair $|\bm{r}_i-\bm{r}_j|$ with using cutoff function of $f_{cut}(\bm{r}_{ij})$. For triplets, the descriptor is in form of a vector $[\bm{r}_{ik}+\bm{r}_{ij}, (\bm{r}_{ik}-\bm{r}_{ij})^2, \bm{r}_{jk}]$ with center atom i, the used cutoff function is $f_{cut}(\bm{r}_{ij})f_{cut}(\bm{r}_{ik})$ \cite{Bartok:2015}.

The parameters used for the pair
descriptor distance\_2b are $cutoff=8.0$, $theta\_uniform=1.0$, $delta=0.5$, and for the triplets descriptor angle\_3b are $cutoff=6.0$, $theta\_fac=0.5$, $delta=0.5$. The nature of these two-body and three-body descriptors makes them perfectly appropriate for the second and third order phonon-phonon interactions.

\section{\label{sec:Results and Discussions}Results and Discussions}

\begin{figure*}
\includegraphics[width=0.95\textwidth]{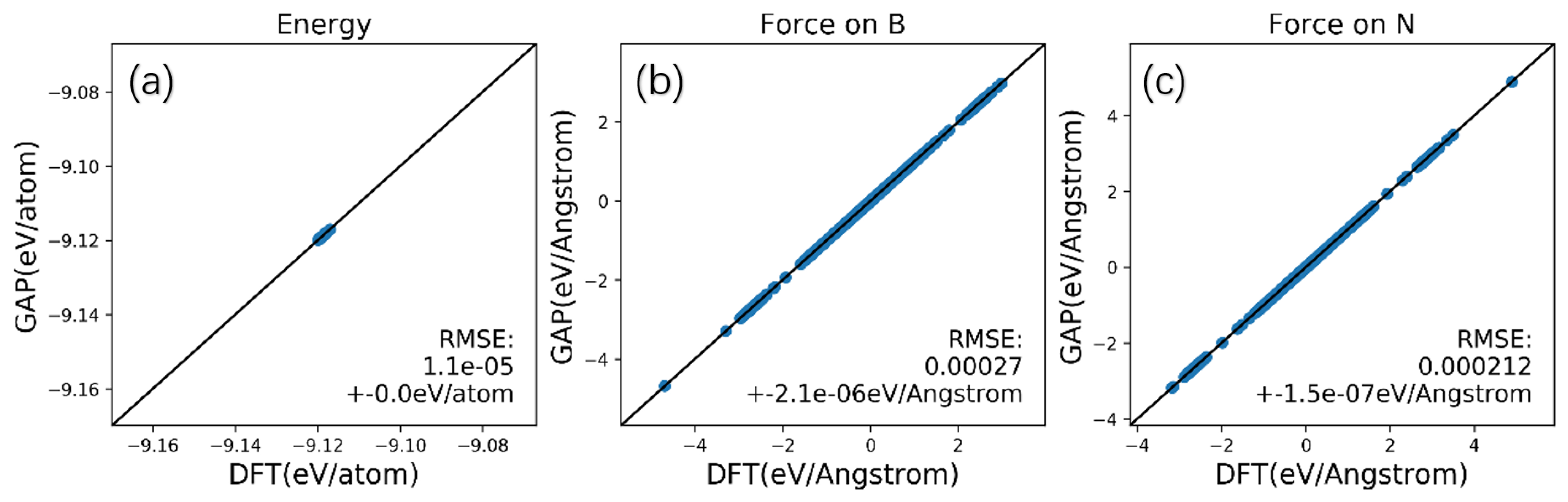}
\caption{\label{fig:Fig1} The comparison of energies and interatomic forces between DFT and MLIP trained using the Manual dataset. (a) the energy; (b) the forces on B atoms; (c) the forces on N atoms.}
\end{figure*}

To verify the construction of MLIP, Figure \ref{fig:Fig1} shows the comparison of total energies and interatomic forces between DFT and GAP results for the test sets, where the training is done using the 325 Manual configurations. The root-mean-squared errors (RMSE) of dataset for energy is $1.1\times10^{-5}$ eV/atom compared with the mean value of energy (-9.12 eV). Whereas the RMSE for forces on the boron (nitrogen) atoms are $0.00027\pm2.1\times10^{-6}$eV/\AA($0.000212\pm1.55\times10^{-7}$eV/\AA), respectively, with the interatomic forces ranging from -4 eV/\AA to 4 eV/\AA. The high accuracy and excellent consistency of the so obtained MLIP gurantee the accuracy of the predicted properties, as discussed later.

\begin{figure}
\centering
\includegraphics[width=1.0\textwidth]{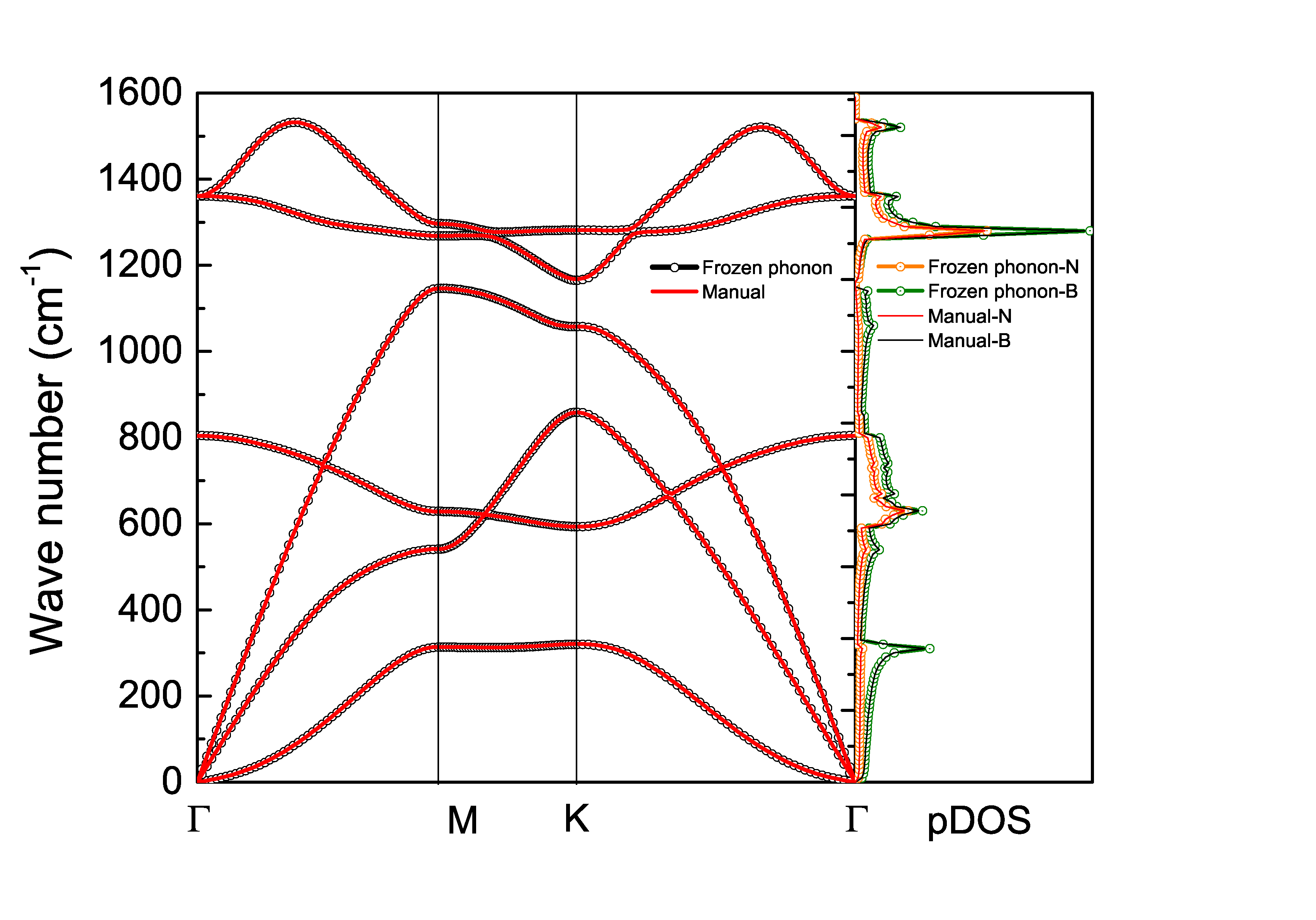}
\caption{\label{fig:Fig2} The comparison for phonon dispersion (left panel) and partial phonon density of states (pDOS, right panel) for frozen-phonon and MLIP based on the Manual training set.}
\end{figure}

Figure \ref{fig:Fig2} shows the phonon dispersion and phonon density of states (DOS) obtained based on the explicit frozen-phonon calculations and MLIP (Manual). It is observed that both the phonon spectra and DOS obtained using MLIP are in excellent consistentcy with the frozen phonon results. That is, MLIP can be applied to get reliable harmonic phonon spectra. Such excellent agreement indicates that both the phonon velocities and distribution function (Eq.\ref{eq:equation1}) can be well reproduced using MLIP.

\begin{figure*}[htp]
\centering
\includegraphics[width=0.48\textwidth]{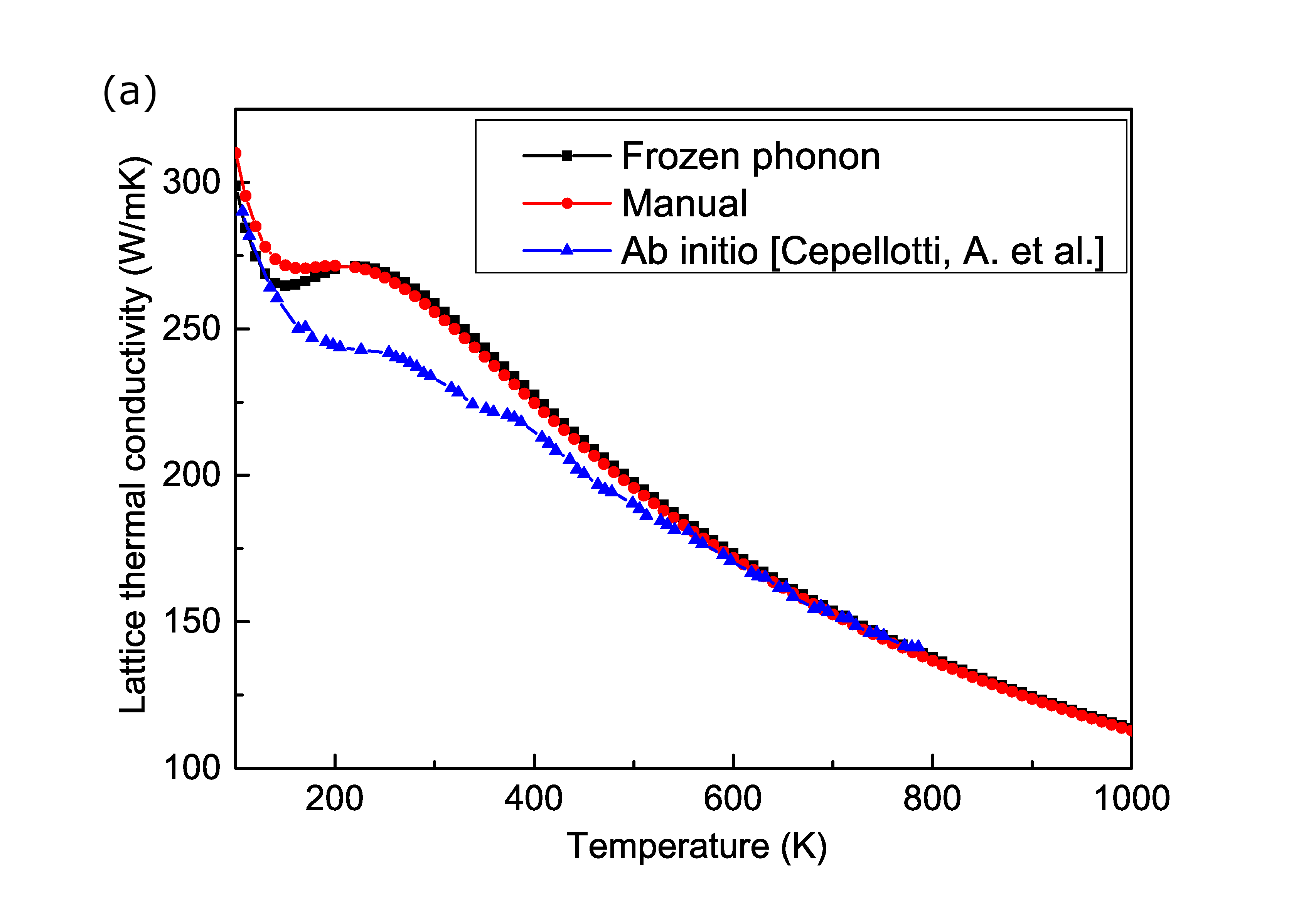}
\includegraphics[width=0.48\textwidth]{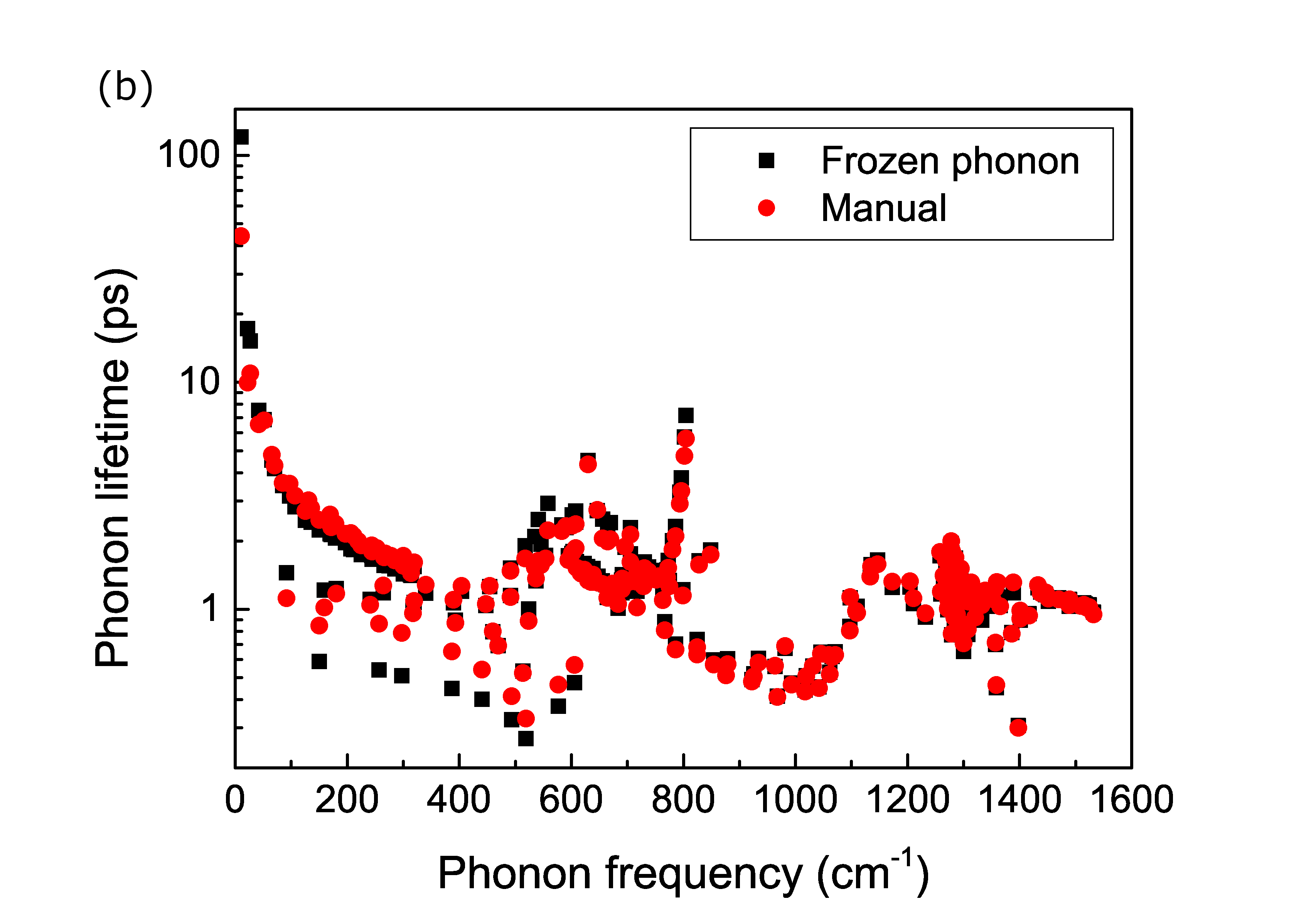}
\caption{\label{fig:Fig3} The comparison of (a) lattice thermal conductivity and (b) phonon lifetime of h-BN monolayers for frozen phonon and MLIP based on the Manual training set. Blue triangles in panel (a) denote the results of Cepllotti {\it et al.}~\cite{Cepellotti:2015}.}
\end{figure*}

Now turn to the thermal conductivity which is originated from the phonon-phonon scattering. According to Eq.\ref{eq:equation1}, the phonon lifetime, which is an essential quantity for the thermal conductivity, can be obtained from the phonon linewidths at irreducible k-points. In order to get the phonon linewidths, the anharmonic terms beyond the quadratic order should be considered, and in this work we considered  only the third-order terms. Figure \ref{fig:Fig3}(a) shows the thermal conductivities in comparison with previous DFT calculations by Cepelloti {\it et al.}~\cite{Cepellotti:2015}, and Figure \ref{fig:Fig3}(b) displays the phonon lifetime obtained using frozen phonon and MLIP.

Obviously, the phonon lifetime obtained using frozen phonon and MLIP is very comparable, especially at high phonon frequency range. This indicates that the MLIP based on the GAP model can account well for the anharmonicity effects. However, at the middle and low frequency range, the phonon lifetime is slightly overestimated in the MLIP results. This is due to the long-range phonon-phonon interactions with low vibrational frequencies, which are not well considered in the training set. Nevertheless, it is fair to say that the MLIP results based on Manual training set can be quantitatively compared with the original frozen phonon results.

Importantly, the thermal conductivities obtained based on frozen phonon and MLIP are in good agreement, as shown in Figure \ref{fig:Fig3}(a). For instance, the deviation of the thermal conductivity obtained using MLIP is as small as -0.73W/mK at 1000K, with a relative error only about 0.71\%. Such a good accuracy persists to the medium temperature range between 200 K and 500 K, {\it e.g.}, the deviation at 400K is 2.50 W/mK, with the relative error of 1.12\%. However, significant deviation is observed below 200K. This can be attributed to the missing long-range phonon-phonon interactions in the MLIP model at the low frequency range as observed in the phonon lifetime (Figure \ref{fig:Fig3}(b)), which plays an important role for thermal conductivity at low temperatures. Last but not least, our frozen phonon results are in good agreement in the medium to high temperature range with those obtained by Cepelloti {\it et al.}~\cite{Cepellotti:2015}, but not for the low temperature below 500 K. The deviation may be originated from the differences in isotopic concentrations and cutoff distance when evaluating the force constants. It is noted that in the low temperature range, the quantum effect should be considered, which is beyond the scope of this work.

Furthermore, in order to verify whether such representative configurations can be selected automatically during the training processes and find out the minimal set, we tested the Incremental method. Five different MLIP models trained with increasing number of configurations were obtained, corresponding to 9.5\% (Incremental-2), 14.1\% (Incremental-3), 18.7\% (Incremental-4), 23.3\% (Incremental-5) and 27.9\% (Incremental-6) of the total configurations, respectively.

Table \ref{tab:table2} shows the RMSE comparison of total energies and atomic forces where Incremental-2, 4, and 6 training-sets are used to construct MLIP. All the models shows good agreement with the DFT results. Note that the Incremental-2 dataset corresponds to only 10\% of the data, suggesting already a good selection of the training set. As the iteration times increase, {\it i.e.}, more configurations are selected as the training set, the RMSE of the resulting interatomic forces decreases. Nevertheless, for the energy prediction, both Incremental-2 and 4 has RMSE energies smaller than that from the Manual set, whereas that of Incremental-6 abnormally increases. It is suspected that this is caused by the fact that specific critical structures are not involved in the training set, so these models are not explicitly enough in predicting the forces. On the other hands, these models also avoid giving huge offsets when facing critical data, so that the RMSE of energy predictions are smaller. 

\begin{table*}[htp]
\caption{\label{tab:table2} RMSE comparison of energies and interatomic forces between DFT and MLIP data trained with (a) Incremental-2, (b) Incremental-4, and (c) Incremental-6 training-sets (cf. Table \ref{tab:table1}).}
\begin{ruledtabular}
\begin{tabular}{lccc}
\textrm{}&
\textrm{Energy (eV/atom)}&
\textrm{Force on B (eV/\AA)}&
\textrm{Force on N (eV/\AA)}\\
\colrule
Incremental-2 & $5e-6$ & $2.82e-4\pm3.1e-7$ & $2.32e-4\pm2.1e-7$\\
Incremental-4 & $6e-6$ & $2.37e-4\pm1.5e-7$ & $2.01e-4\pm1.0e-7$\\
Incremental-6 & $1e-5$ & $2.32e-4\pm1.4e-7$ & $1.96e-4\pm9.0e-8$\\
\end{tabular}
\end{ruledtabular}
\end{table*}

\begin{figure}
\centering
\includegraphics[width=1.0\textwidth]{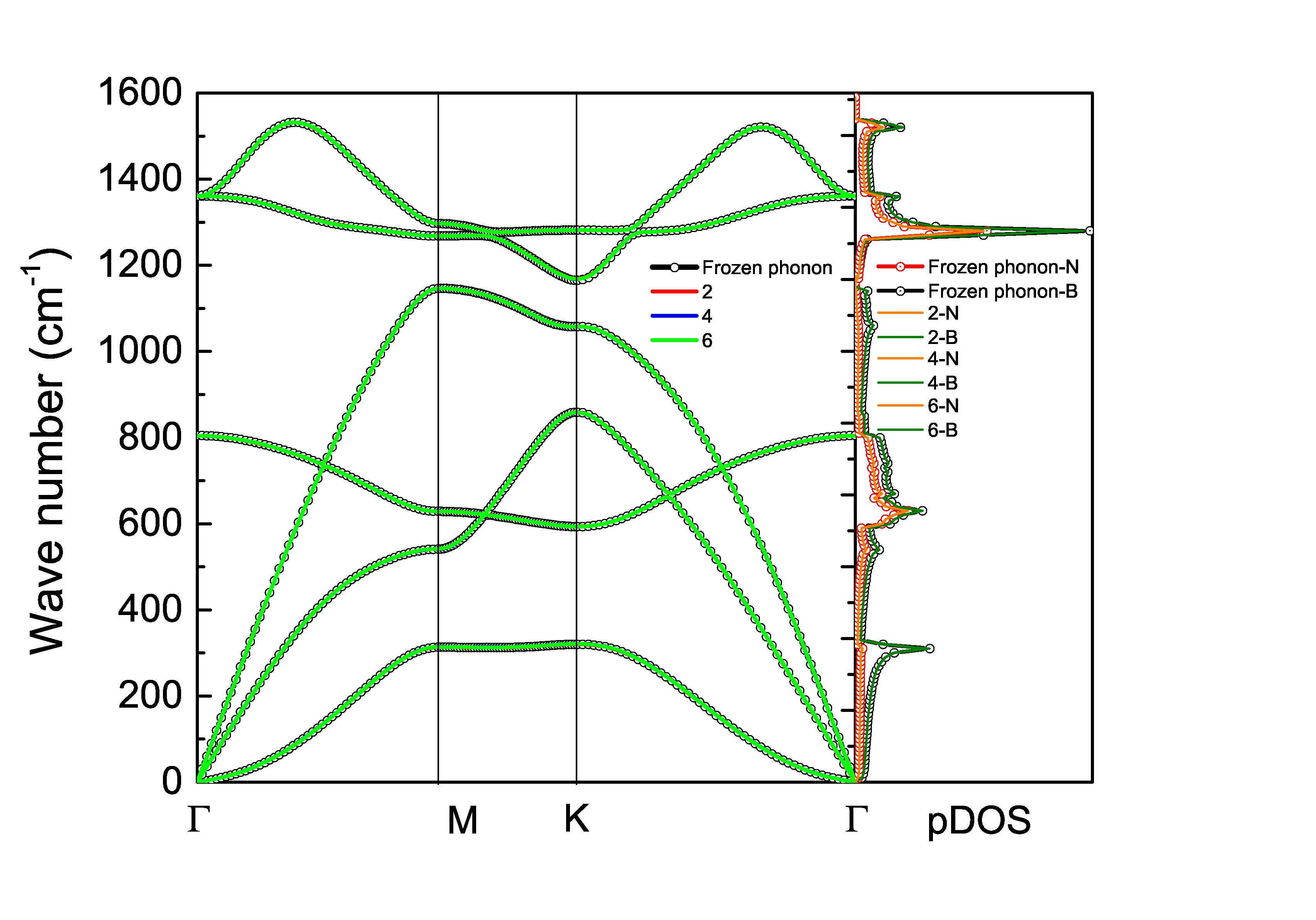}
\caption{\label{fig:Fig5} The comparison for phonon dispersion and partial phonon density of states obtained using frozen phonon and MLIP based on Incremental-2, 4 and 6 training sets.}
\end{figure}

Although the RMSEs of the Incremental models are slightly deviated from each other, all the models give rise to accurate phonon spectra as shown in Figure \ref{fig:Fig5}. In this regard, a small number of well-selected configurations are enough to reproduce the harmonic properties.

\begin{figure*}[htp]
\centering
\includegraphics[width=0.48\textwidth]{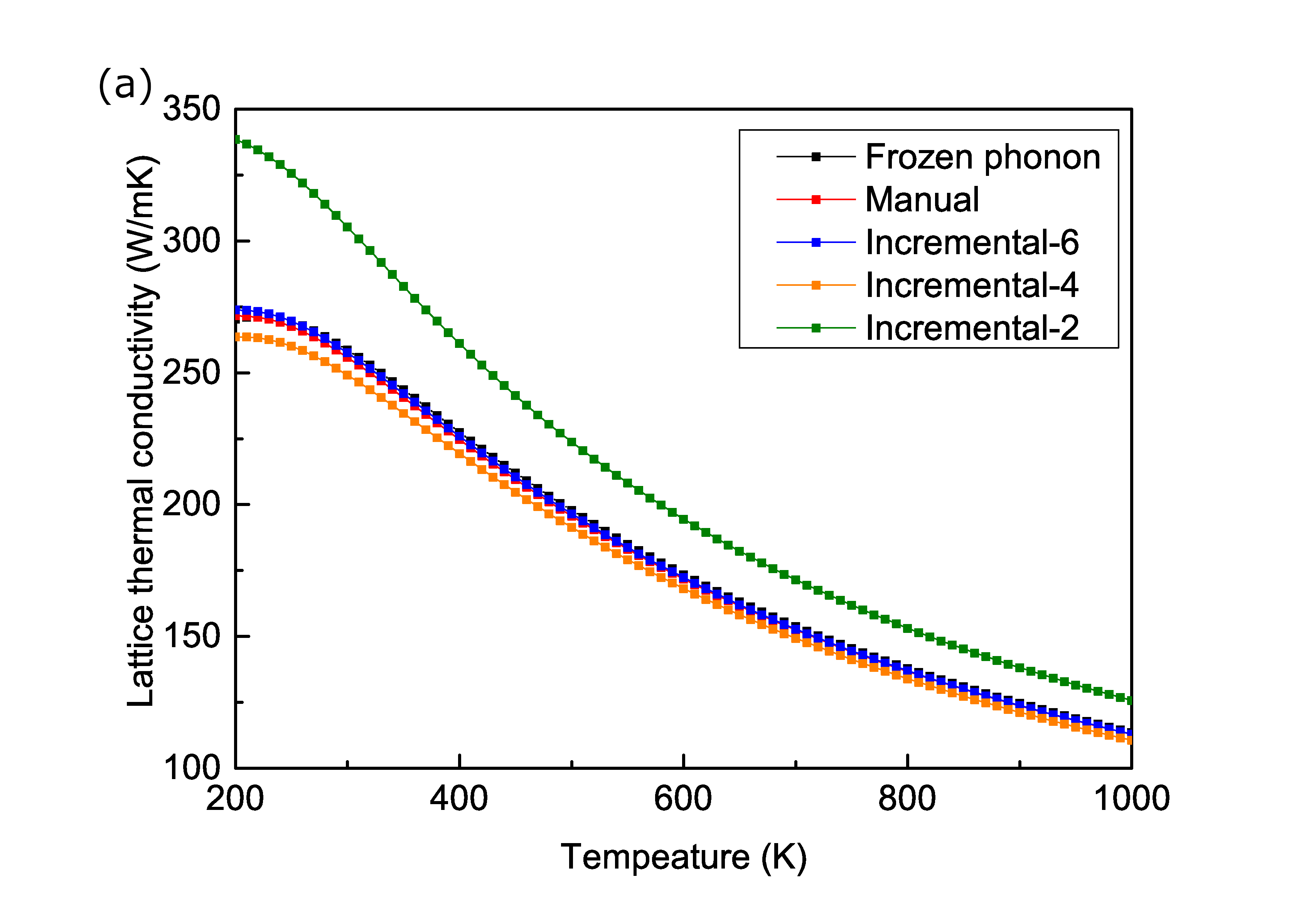}
\includegraphics[width=0.48\textwidth]{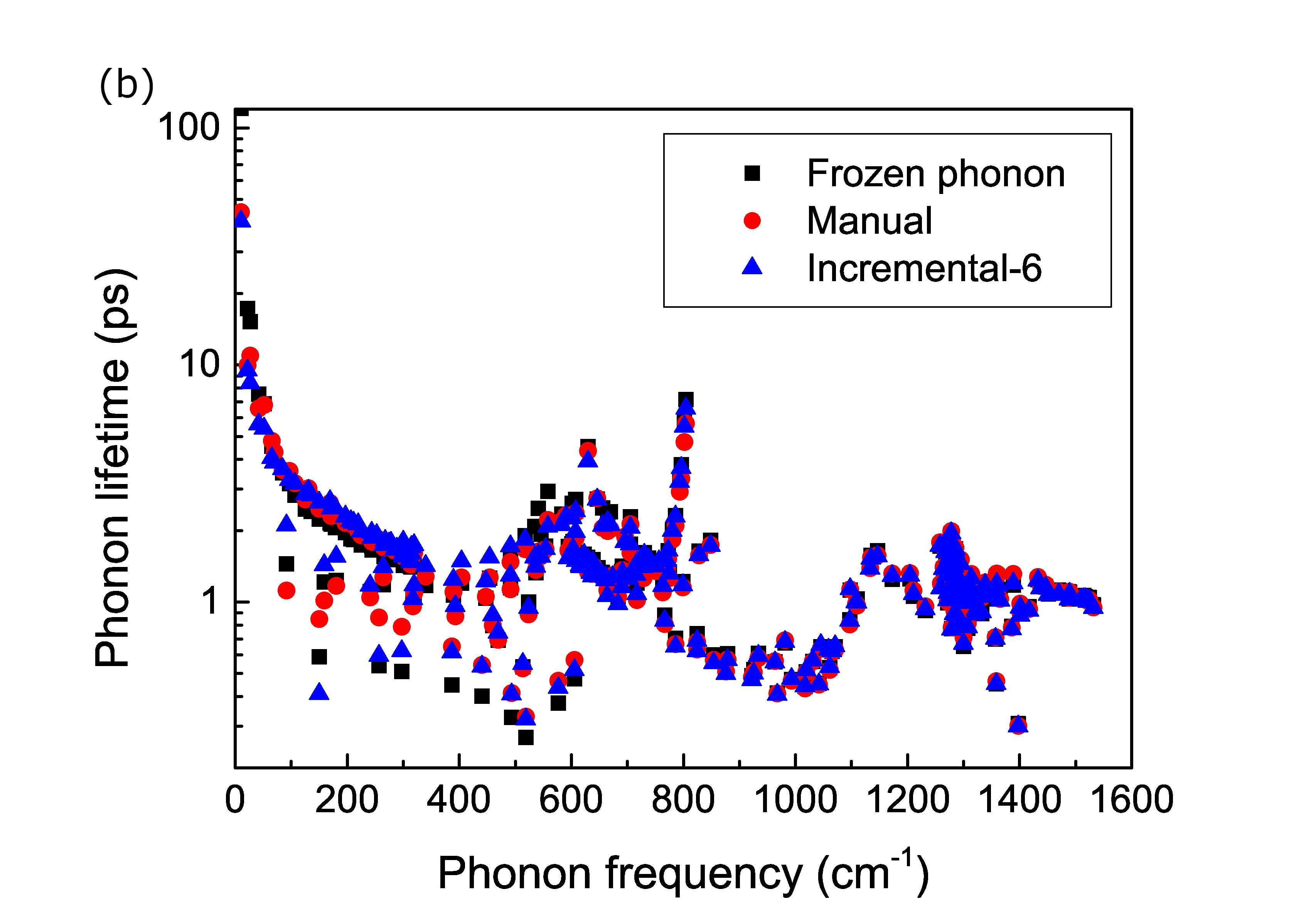}
\caption{\label{fig:Fig6} The comparison of: (a) lattice thermal conductivity of frozen phonon and MLIP based on the Manual and Incremental-2/4/6 training-sets; (b) phonon life time of frozen phonon and MLIP based on Manual and Incremental-6 training-sets.}
\end{figure*}

Figure \ref{fig:Fig6}(a) shows the resulting thermal conductivities obtained using different methods as specific above, the plot of all methods is presented in Figure S3. Apparently, the more configurations used in the training set, the more accurate the resulting thermal conductivities, in comparison with the frozen phonon results. Although all Incremental models give rise to excellent harmonicitic lattice dynamical properties (Figure \ref{fig:Fig5}), the thermal conductivities based on most Incremental models are overestimated, except for the Incremental-4 case with 18.7\% of the configurations where the thermal conductivity is slightly underestimated (Table \ref{tab:table2} Incremental-4). This suggests that the anharmonic terms, {\it i.e.}, the phonon-phonon interactions, requires higher accuracy of MLIP. Furthermore, a clear trend of convergence is observed (Figure \ref{fig:Fig6}(a)), and the result from the Incremental-6 case (with 27.9\% configurations) are in good agreement with the frozen phonon result (Figure \ref{fig:Fig6}(a)), 
{\it e.g.}, the deviation of the Incremental-6 result is about -0.648 W/mK at 1000K, with a relative error about 0.63\%. This is even slightly better than that based on the MLIP obtained using the Manual training set. This can be understood based on the phonon lifetime shown in Figure \ref{fig:Fig6}(b), where the MLIP obtained with Incremental-6 training-set can reproduce well the major features of the phonon lifetime. In this regard, the computational time for lattice thermal conductivities can be significantly reduced by selecting the most representative configurations as the training set. It is admitted that currently we have evaluated the forces for all the configurations based on DFT calculations, which are used to benchmark the MLIP trained using part of the configurations. In order to be able to implement such a scheme without doing calculations on the whole set, we suspect that active learning with uncertainty quantification can be applied so that only the most representative configurations will be selected and calculated explicitly based on DFT~\cite{Frederiksen:2004}. 

\quad
\section{\label{sec:Summary}Summary}
In conclusion, GAP-type MLIPs constructed using different training sets are applied to evaluate the lattice dynamics and thermal conductivities of h-BN monolayers. It is illustrated that the phonon spectra can be accurately obtained using a small set (about 30\%) 
of configurations, leading to an efficient way of performing calculations on the lattice thermal conductivities. It is further observed that for three-dimensional materials like beta-Sn (not shown), even less than 30\% configurations can be selected automatically and used to construct a MLIP to evaluate the thermal conductivities accurately. Moreover, the thermal conductivity calculations require more accurate construction of MLIPs than the harmonic lattice properties, which can be applied to benchmark the quality of MLIPs.

It is suspected that active learning with proper uncertainty quantification can be applied to select the most representative configurations as the training set, so that the numerical efforts to evaluate the thermal conductivity can be significantly reduced, which will be saved for future studies. And it is also suspected that such a workflow is capable in constructing MLIPs for MD calculations as well, while proper configurations related to desired properties should  be  generated  and  considered  to  construct the more suitable MLIPs.

\begin{acknowledgments}
Yixuan Zhang thanks the financial support from Fullbright-Cottrell Foundation. Hongbin Zhang acknowledges the financial support from the Deutsche Forschungsgemeinschaft (DFG, German Research Foundation) Project-ID 405553726–TRR 270. Teng Long thanks the financial support from China Scholarship Council. Calculations of this research were conducted on the Lichtenberg high performance computer of TU Darmstadt.
\end{acknowledgments}

\end{document}